\documentstyle[aps,prb]{revtex}
\catcode`\"=\active \let"=\" \let\3=\ss
\renewcommand{\baselinestretch}{2}
\tightenlines
\renewcommand{\d}{\displaystyle}
\def\fcap#1{\rightskip1cm\noindent\small #1\normalsize\rightskip0cm}
\newcommand{\Kr}[1]{\left( #1\right)}
\newcommand{\Ke}[1]{\left[ #1\right]}
\newcommand{\Kg}[1]{\left\{ #1\right\}}

\newcommand{\ex}[1]{{\rm e}^{{\rm i}#1}}
\newcommand{\ve}{\varepsilon}
\newcommand{\vf}{v_{\rm F}}
\newcommand{\kf}{k_{\rm F}}
\begin{document}
\title{Tomonaga--Luttinger parameters for quantum wires}

\author{Wolfgang H"ausler$^{1,2,3}$, Lars Kecke$^2$, and A.~H.~MacDonald$^{1,4}$}

\address{$^1$ Department of Physics, Indiana University, 701 E Third Street,
Swain Hall--West 117, Bloomington, IN 47405, U.S.A.\\
$^2$ I.~Institut f"ur Theoretische Physik der Universit"at Hamburg,
Jungiusstr.~9, D-20355 Hamburg, Germany\\
$^3$ Fakult"at f"ur Physik, Albert-Ludwigs-Universit"at Freiburg,
D-79104 Freiburg, Germany\\
$^4$ Department of Physics, University of Texas at Austin, Austin TX, 78712,
U.S.A.}
\maketitle

\begin{abstract}

The low-energy properties of a homogeneous one-dimensional
electron system are completely specified by two
Tomonaga-Luttinger parameters $K_{\rho}$ and $v_{\sigma}$. In
this paper we discuss microscopic estimates of the values of
these parameters in semiconductor quantum wires that exploit
their relationship to thermodynamic properties of the quantum
wire. Motivated by the recognized similarity between
correlations in the ground state of a one-dimensional electron
liquid and correlations in a Wigner crystal, we evaluate these
thermodynamic quantities in a self-consistent Hartree-Fock
approximation. According to our calculations, the Hartree-Fock
approximation ground state is a Wigner crystal at all electron
densities and has antiferromagnetic order that gradually evolves
from spin-density-wave to localized in character as the density
is lowered. Our results for $K_{\rho}$ are in good agreement
with weak-coupling perturbative estimates $K_{\rho}^{\rm pert}$
at high densities, but deviate strongly at low densities,
especially when the electron-electron interaction is screened at
long distances. $K_{\rho}^{\rm pert}\sim n^{1/2}$ vanishes at
small carrier density $n$ whereas we conjecture that
$K_{\rho}\to 1/2$ when $n\to 0$, implying that $K_{\rho}$ should
pass through a minimum at an intermediate density. Observation
of such a non-monotonic dependence on particle density would
allow to measure the range of the microscopic interaction to be
determined. In the spin sector we find that the spin velocity
decreases with increasing interaction strength or decreasing
$n$. Strong correlation effects make it difficult to obtain
fully consistent estimates of $v_{\sigma}$ from Hartree-Fock
calculations. We conjecture that $v_{\sigma}/\vf\propto n/V_0$
in the limit $n\to 0$ where $V_0$ is the interaction strength.

\end{abstract}

\maketitle

\noindent PACS~: 71.10.Pm, 71.10.-w, 71.10.Hf, 71.45.Lr

%\twocolumn
\section{Introduction and overview}
It has been known for some time that one--dimensional (1D)
metals are different from their higher dimensional Fermi-liquid
cousins \cite{tomonaga,luttinger}. It is generally believed that at
low energies and long wavelengths one dimensional electron systems
can, under very general circumstances, be
described as `Tomonaga--Luttinger (TL) liquids \cite{haldane},
although it has nearly always been difficult to provide incontrovertible
experimental evidence. Interest in TL liquids has been heightened
in recent years by new physical realizations, including quantum
Hall edge systems \cite{wen,kanefisher}, carbon nanotubes
\cite{cnts,tubes}, and semiconductor quantum wires
\cite{tarucha,yacoby} in particular. Like Fermi liquid theory,
TL theory can be used to relate low-temperature, low-frequency,
long-wavelength properties to a small number of parameters in which the
microscopic physics of particular systems is encoded.
For example, TL theory predicts that for continuous one--channel
quantum wires the quantized conductance is renormalized by the factor
\cite{apel} $\:K_{\rho}\:$ at frequencies larger than
\cite{reservoirs} $\:\vf/L\:$ ($\:L\:$ is the wire length and
$\:\vf\:$ the Fermi velocity). Surprisingly, low-energy orthogonality
catastrophes lead to spectral functions that follow power
laws\cite{meden,kanefisher}
specified in terms of the same parameter. In many cases
(up to logarithmically slowly varying prefactors\cite{finkelshtein,agsz}
associated with the presence of marginal operators like back
scattering in spin sector) non-universal power laws specified by TL theory
parameters are also predicted for the behavior of correlation functions
at distances much larger than the spatial range of interactions.
(The strictly infinite range Coulomb interaction case requires special
considerations.\cite{schulz93,gogolin94a}). Microscopic theory still has an
important role at low energies however, in estimating the values of these
parameters. This is especially important because the
distinction between Fermi liquids and Luttinger liquids on the
basis of a set of experimental data over a limited temperature
or energy range is sometimes subtle, and the range of energies
over which TL behavior is expected is often not accurately
known. Approximate values of expected TL parameters can play a
role in determining whether or not an experimental result
reflects TL behavior\cite{yacoby}. In addition, as this approximate
calculation shows, the problem of understanding the value of the
two independent TL parameters of a homogeneous one-dimensional
electron system is a challenging many-body problem that is
interesting in its how right.

Four TL parameters characterize the low energy properties of
interacting spinful electrons moving in one channel. For the
charge ($\:\nu=\rho\:$) or spin ($\:\nu=\sigma\:$) sector, the
parameter $\:K_{\nu}\:$ fixes the exponents for most of the power
laws and $\:v_{\nu}\:$ is the velocity of the long wave length
excitations. Symmetries in the charge or in
the spin sector reduce the number of independent parameters in the
case of a one-dimensional electron gas system~:
spin rotation invariance enforces \cite{voit} $\:K_{\sigma}=1\:$
while Galilean invariance implies that \cite{circcurr}
$\:v_{\rho}=\vf/K_{\rho}\:$. The later identity does not apply,
for example, in lattice models since it requires continuous
translational invariance; in that case $\:v_{\rho}\:$ and
$\:K_\rho\:$ must be determined independently. This leaves
$\:K_{\rho}\:$ and $\:v_{\sigma}\:$ as the only two independent
TL parameters for single-channel semiconductor quantum wires
since they can be accurately described by a continuum envelope
function approximation.

In Fermi liquids a traditional and successful strategy has
separated the phenomenological application of Fermi liquid
theory from the microscopic evaluation of its parameters. To
date most theoretical TL activity has focused on
phenomenological applications; confident interpretation of
experiments will require reliable microscopic estimates of the
theory's parameters for the various physical systems of current
interest. The evaluation of Fermi liquid parameters in two and
three dimensional metals is one of the classic early topics in
many electron physics, with considerable recent progress coming
from quantum Monte Carlo calculations \cite{ceperley}. Still,
useful physical insight and reasonable accuracy has resulted
from less computationally cumbersome approaches. In this paper
we discuss what can and cannot be learned about the values of
TL parameters in semiconductor quantum wires and the physics of
their dependencies on system geometry using unrestricted
Hartree-Fock estimates of ground state energies. The
Hartree-Fock approximation can yield very reliable estimates to
the boundary exponents \cite{schonhammer} describing tunneling
into the end of quantum wires and, as a microscopic approach,
gives information about quantities not reliably accessible in
the TL formalism, including absolute values for the prefactors
of power laws.

For non-interacting electrons the TL parameter
$\:K_{\rho}=1\:$. With repulsive interactions its value should
decrease and go to zero in the limit of very strong or long
ranging \cite{gogolin94a} interactions. For microscopic interaction
potential $\:V(x-x')\:$ the formula
\begin{equation}\label{pop}
K_{\rho}^{-1}=\sqrt{1+4mV(k=0)/\pi^2n}
\end{equation}
is commonly used in the literature. It depends on the
carrier density $n$, the effective mass $m$ and the
$k=0$ Fourier component of the interaction
\begin{equation}\label{vfour}
V(k)=\int{\rm d}x\;V(x)\cos kx\;.
\end{equation}
Relation (\ref{pop}) can be motivated by lowest order
perturbation theory, or by the random phase approximation
\cite{dassarma}, though it misses the Fock contribution for
spinful electrons in 1D. Any naive higher order perturbative
contribution is
divergent, only the infinite subsums that are conveniently
captured using a perturbative renormalization approach are
finite \cite{larkin}). Eq.~(\ref{pop}) completely ignores the
renormalizing influence of short wave length modes in
determining the actual values of the effective interaction.
Higher order perturbative renormalization group calculations
demonstrate how the interaction parameters are coupled and
renormalize as short wave length contributions are integrated
out \cite{solyom}. One important example is back scattering,
across the Fermi line, of opposite spins Fermions, the so called
$g_1$--process, that spoils the separate conservation of the
number of left and right moving particles of a given spin and
therewith is not included in the TL model. This interaction,
which is finite in leading order perturbation theory, scales to
zero during renormalization, restoring SU(2) spin rotation
symmetry and the validity of the TL at low energies
\cite{luther}. Even for a model of spinless electrons, the
parameters will rescale at low energies, reflecting other
irrelevant operators that are omitted in the TL model such as
non-linear dispersion of the kinetic energy in the microscopic
Hamiltonian \cite{haldane}. As a consequence Eq.~(\ref{pop})
cannot be used to estimate $K_{\rho}$ when interactions are
strong.

How interactions influence the spin sector is even less certain.
According to textbook knowledge \cite{mahan} the spin velocity
would not be altered by interaction forces that act only on
spatial coordinates and thus not in the spin sector. Other work
\cite{schulz93} includes exchange contributions in the Bose form
of the Hamiltonian in a way violating the SU(2) invariance
property, $\:K_{\sigma}=1\:$. On the other hand, changes in
$\:v_{\sigma}/\vf\:$ are quite crucial to various physical
properties. It influences for instance the magnetic
susceptibility, the $g$--factor, and spin transport properties.
The latter are particularly important for potential
one-dimensional spintronic devices \cite{spintransistor}. In
one-dimensional channels \cite{tsukagoshi99} for example the
spin conductance \cite{balents}, and Rashba precession in the
presence of spin orbit coupling \cite{whrashba}, depend on
$\:v_{\sigma}\:$. Most directly $\:v_{\sigma}\:$ can be
measured by inelastic Raman scattering in the `depolarized'
configuration with perpendicularly polarized incident and
outgoing light \cite{abstreiter,goni}.

To date relations between the microscopic electron--electron
interaction and resulting TL parameters have been established
for models of primarily theoretical interest, such as the Kondo
Lattice Model \cite{shibata}, the Hubbard model \cite{hubbard},
and the $\:t-J$--model.\cite{kawakami} For the latter two
models the ground state energies are known exactly, either
analytically in certain limiting cases or by solving the
Bethe--Ansatz equations numerically. For these repulsive short
range interaction models $\:K_{\rho}\:$ is found to be
confined to the range $\:1/2\le K_{\rho}\le 1\:$ and it has
been argued \cite{hubbard} in the limits of either infinite
interaction strength or vanishing particle density are
equivalent to non-interacting spinless Fermions with $\kf$ being
replaced by $2\kf$ so that $\:K_{\rho}\to 1/2\:$ in either of
these limits in order to recover the correct asymptotic decay of
the density-density correlation function.
For the $\:t-J$--model, TL parameters away from the
super-symmetric point ($\:J/t=2\:$) have been obtained by using
ground state energies from exact diagonalization calculations
\cite{ogata}. The Sutherland model for spinless Fermions, where
$\:V(x)=\lambda/x^2\:$, has proven to be a TL at low energies
\cite{kawakami2}. The asymptotic decay of its one-particle
Greens function implies that
$\:K_{\rho}=2/(1+\sqrt{1+2\lambda})\:$ with $\:1\ge K_{\rho}\ge
0\:$ for repulsive interactions, $\:\lambda>0\:$. The
compressibility of this system is proportional to
$\:K_{\rho}^2\:$ and satisfies Eq.~(\ref{compress}) below. For
quantum wires with long but finite range Coulomb interactions
the values of the TL parameters have been determined previously
by extensive quantum Monte Carlo calculations \cite{qmc}.
However, the limits on the number of particles and lattice
points in real space for which these calculations can be carried
out in a reasonable time places limits on the range of particle
densities over which accurate Monte Carlo results can be
obtained. In particular the low-density regime where
interactions are strongest is difficult to reach. In this work
we also exploit the thermodynamic relations between the uniform
static compressibility $\:\kappa\:$ and the TL parameter in the
charge sector. For quantum wires we have (cf.\ Refs.\
[\onlinecite{haldane,hubbard}])
\begin{equation}\label{compress}
\frac{1}{\kappa}\equiv\frac{\partial^2(E_0/L)}{\partial n^2}=
\frac{\pi}{2}\frac{v_{\rho}}{K_{\rho}}=
\frac{\pi}{2}\frac{v_{\rm F}}{K_{\rho}^2}\quad,
\end{equation}
where the last equality uses Galilean invariance.

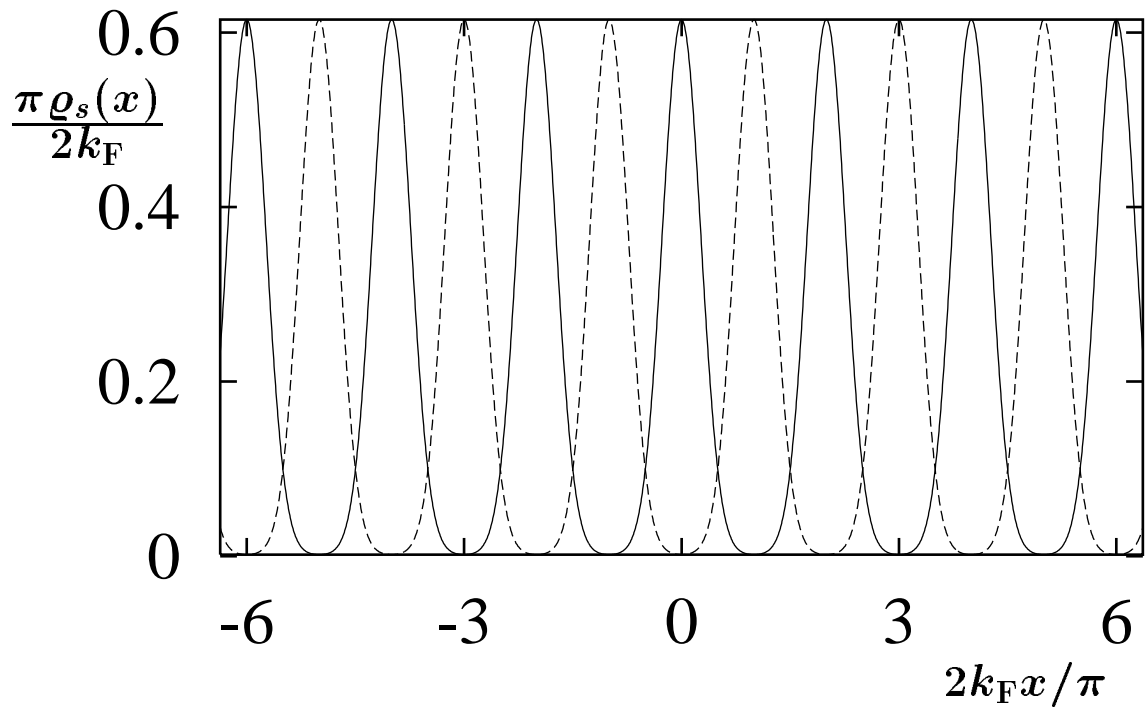
\begin{figure}[h]\caption{\label{densities}}
\fcap{Figure~1: Charge densities $\:n_{\uparrow}(x)\:$ (solid)
and $\:n_{\downarrow}(x)\:$ (dashed) as a function of position
along the wire $\:x\:$ in units of the mean electron spacing for
$\:\kf d=0.15\:$ and $\:R/d=5.66\:$. We argue that these charge
densities in the broken symmetry Hartree-Fock states are a good
approximation to typical configurations in the fluctuating
one-dimensional electron liquid that does not have broken
translational symmetry.}
\end{figure}

Density--density correlation function calculations performed
within the TL model have been used \cite{schulz93} to show that
long range interactions $\:V(|x|)\sim|x|^{-a}\:$ put the
one-dimensional Fermion system ground state into a Wigner
crystal when $\:a<1\:$, irrespective of the strength of $\:V\:$.
The Coulomb case, $\:a=1\:$, is marginal and density--density
correlations decay extremely slowly, slower than any power law
at large distances. This observation suggests that the ground
state energy $E_0$ and therefore any $T=0$ thermodynamic
property should be accurately estimated by the unrestricted
Hartree-Fock approximation, for which the ground state {\em is}
a Wigner crystal. At high densities it is known \cite{mahan}
that the Hartree-Fock (HF) approximation reproduces the leading
order perturbative RG result for $K_{\rho}$. Since it
spontaneously breaks translational symmetry, the HF
approximation correlation functions have infinite range. The
tails of the correlation functions are therefore given slightly
incorrectly. Much more important for the energy however, is the
accurate estimate of the magnitude of the short-distance
correlation function oscillations, illustrated in
Fig.~\ref{densities}. The dominant periods for charge and spin
densities are in agreement with TL model calculations, which are
however, not able to estimate the amplitude of the oscillations.
In the real correlation function, these oscillations are
multiplied by an envelope function whose decay properties are
inaccessible to Hartree-Fock theory but can be calculated from
the TL theory. The accuracy of self-consistent Hartree-Fock
energy estimates, particularly at low-densities, has been
previously been established in other interacting electron
systems.\cite{yoshioka}

We compare our results for ground state energies and
compressibilities $\kappa$ with estimates obtained within
the harmonic approximation to the classical Wigner crystal, cf.\
(\ref{wigqufl},\ref{classic}) and within perturbation theory
\cite{pertmean}. The perturbative expression (\ref{pert}) we
use (see below) for the charge sector TL parameter
$\:K_{\rho}\:$ turns out to be surprisingly accurate over a wide
range of carrier densities including typical experimental ones.
Only below densities corresponding to values $\:r_{\rm s}>1.3\:$
of the usual $r_{\rm s}$--parameter used to measure the
interaction strength in metals (see below), do we find {\em
smaller} $\kappa$ in the selfconsistent HF solution than given
by Eq.~(\ref{pert}) and (\ref{compress}). At even lower
densities $\:n\lesssim 1/R\:$, where $R$ is the long but finite
range discussed below that we use for the electron-electron
interaction, Eq.~(\ref{pert}) predicts that $\kappa$ goes to
zero whereas HF theory yields a finite limiting value,
after $\kappa$ has passed through a minimum.
This minimum is also reproduced by the harmonic approximation.
In the latter approximation, however, $\kappa$ diverges as
$\:n\to 0\:$. We shall give arguments supporting the conjecture
that the true $\:K_{\rho}\:$ stays finite as $\:n\to 0\:$ and
indeed that $\:K_{\rho}\to 1/2\:$ in this limit.

We also analyze the spin sector and compare different approaches
in the attempt to determine the spin velocity $\:v_{\sigma}\:$.
The simplest estimate is again low order perturbation theory for
the magnetic susceptibility. Other estimates can be obtained by
starting with the assumption that the system is close to an
antiferromagnetic Heisenberg spin chain at low densities. It
turns out that the estimates for $\:v_{\sigma}\:$ that follow
from different plausible schemes differ substantially.
Furthermore, at smaller particle densities they are not in good
agreement with earlier QMC estimate \cite{qmc}. Although we
are able to conclude that $\:v_{\sigma}\:$ is small at
low-densities, our results for the spin sector do not
substantially improve on existing Monte Carlo results.

\section{Microscopic Interaction}
Now we discuss the form of realistic interactions $V$ along the
$x$--direction of a quantum wire. At inter--particle separations
$\:|x-x'|\:$ larger than the diameter $d$ of the wire
$\:V(|x-x'|\gtrsim d)=e^2/\epsilon|x-x'|\:$ will be of the
Coulomb form, irrespective of the detailed shape of the
transversal potential. If the material enclosing the wire is
insulating with dielectric constant $\epsilon'$, the Coulomb form
still holds at larger distances, but with dielectric $\epsilon$
replaced by \cite{dietl} $\epsilon'$. We assume here equal
dielectric constants $\:\epsilon \sim \epsilon'\:$, the case that
applies to gated \cite{tarucha} as well as in cleaved edge overgrowth
structures \cite{yacoby}. Eventually, at an inter--particle
separation exceeding the distance $\:R\:$ to the closest metallic
system that is extended in the direction along the wire. This metallic
screening can be supplied by carriers in nearby metallic gates,
including those used to define the quantum wire. Assuming that
the screening plane and the quantum wire are parallel
$\:V(|x-x'|>R)\sim 1/|x-x'|^3\:$ because of the formation of
dipoles from image charges. The interaction (\ref{interaction})
below accounts for this cut-off at large particle separations.

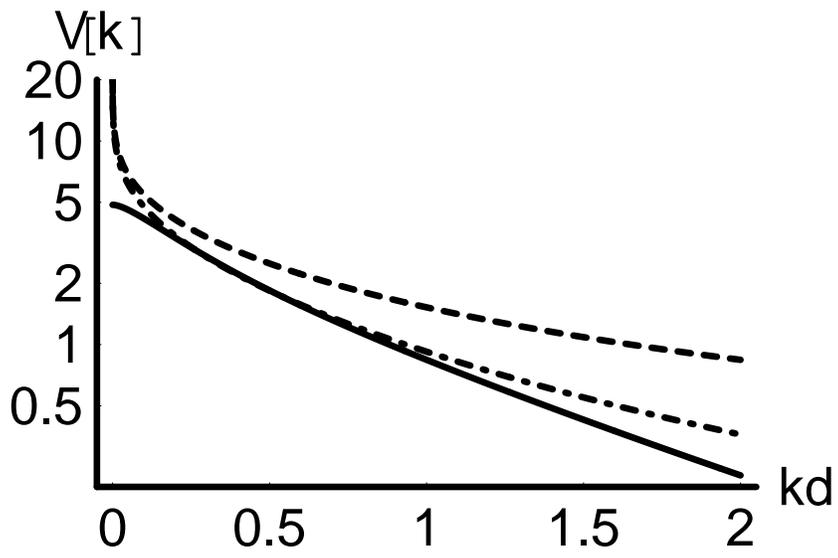
\begin{figure}[h]\caption{\label{potentials}}
\fcap{Figure~2: Fourier transforms of three different forms of
microscopic electron--electron interactions, $\hat{V}(k)=V(k)/\vf$,
as described in the text. Solid: the form (\ref{interaction}) we
use in the present work, dashed: the 2D heterostructures model,
dash dotted: the 3D cylindrical case intended for cleaved edge
overgrowth systems}
\end{figure}

At distances $\:|x|\lesssim d\:$ shorter than the wire width,
the precise transverse form of the electronic wave function
influences $\:V(|x|)\:$. For example, in 2D hetero--structures
all electrons share a common growth direction wave function. If
this wave function is approximated by a harmonic oscillator with
characteristic length $d$, and thickness in the 2nd confined
direction is neglected, we have\cite{huoconnell} $\:V_{\mbox{\tiny
2D}}(x)=(e^2/2\sqrt{\pi}\epsilon d){\rm e}^{x^2/8d^2}
K_0(x^2/8d^2)\:$ where $K_0$ denotes a Bessel function. It is
more realistic to include finite thickness in both confined
directions. For example, for 3D wires with circular cross
sections, a model which might be appropriate for
cleaved-edge-overgrowth systems, we have\cite{friesen}
$\:V_{\mbox{\tiny 3D}}(x)=(e^2/\sqrt{2/\pi}\epsilon d){\rm
e}^{x^2/2d^2}\mbox{erfc}(x/\sqrt{2}d)\:$ again using harmonic
oscillator ground states of widths $\:d/\sqrt{2}\:$, now for
both of the transverse directions. Neither of these forms is
smooth at $\:x=0\:$, an artifact of assuming factorized wave
functions. At short distances, corresponding to high energies,
the factorization assumption needs to be refined. It leads to
the unphysically slow decay of the Fourier transforms
$\:\hat{V}_{\mbox{\tiny 2D}}(k)=(e^2/\epsilon\vf){\rm
e}^{k^2d^2/2}K_0(k^2d^2/2)\:$, and $\:\hat{V}_{\mbox{\tiny
3D}}(k)=(e^2/\epsilon\vf){\rm e}^{k^2d^2/2}E_1(k^2d^2/2)\:$, as
seen in Figure~\ref{potentials}, $E_1$ is the exponential
integral. Also included in Figure~\ref{potentials} is the more
realistic form,
\begin{equation}\label{interaction}
\hat{V}(k)=\frac{2}{a_{\rm B}\kf}\Ke{K_0(kd)-K_0(k\sqrt{d^2+4R^2})},
\end{equation}
which accounts for the image potential term from a remote
screening plane separated by $R\gg d$. Its real space form
is displayed in Eq.\ (\ref{xinteraction}). This interaction remains
finite for $k \to 0$ and decreases more quickly for large
$\:k\:$ and will be used in the present work. Here, and in the
following we measure the interaction in units of the Fermi
velocity so that $\hat{V}$ is dimensionless in
Eq.~(\ref{interaction}). Its strength, in comparison with the
kinetic energy, scales with the dimensionless parameter
$\:r_{\rm s}:=1/(2na_{\rm B})=\pi/(4\kf a_{\rm B})\:$ depending
on density $\:n\:$ and the effective Bohr radius $\:a_{\rm
B}\:$. In our calculations we assume $\:a_{\rm B}=2d\:$.

Thus, two parameters
\begin{equation}\label{parameters}
R/d\qquad\mbox{and}\qquad\kf d
\end{equation}
characterize the range and the strength of our model interaction
(\ref{interaction}), respectively. They both can be extracted
quite reliably from experiment, $R$ from the sample lay out and
$d$ from the energy $\sim \:1/md^2\:$ of inter-subband excitations.
Typical distances to metallic gates and typical wire width, as
reported eg.\ in [\onlinecite{tarucha}] and
[\onlinecite{yacoby}], correspond to values for $\:R/d\:$ ranging
from 5 to 14. Typical single wall
carbon nanotube systems, on the other hand, would correspond to much
larger $\:R/d\:$ values because of their extremely small diameters.
Many of our calculations are for
$\:R/d=5.66\:$ or 35.36. Note also that electron densities
should be sufficiently low ($\:\kf d<\sqrt{2}\:$ within the
parabolic approximation for the transverse confinement) to
prevent occupation of the second subband.

\section{Ground State Energy}\label{gse}
According to (\ref{compress}) we need to calculate ground state
energies for different particle densities. In this work we
employ the unrestricted Hartree-Fock approximation where
details are described in Appendix~\ref{hfthy}. Results of these
calculations are included in subsequent Figures.

The close proximity to the Wigner crystal and the Bose character
of all of the low energy excitations suggests comparing with
the ground state energy density of the harmonic crystal in 1D,
\begin{equation}\label{wigqufl}
E_0^{\rm wc}=E_0^{\rm classical}+\frac{1}{2}\int_{-\kf}^{\kf}
\frac{{\rm d}k}{2\pi}\;\omega(k)\quad.
\end{equation}
Here,
\begin{equation}\label{classic}
E_0^{\rm classical}=\frac{1}{2L}\sum_{i\ne j}V(|i-j|\pi/2\kf)
\end{equation}
is the classical contribution and the zero point energy follows
from the phonon dispersion
\begin{equation}\label{qufl}
\omega^2(k)=\frac{1}{m}\sum_{j=1}^{\infty}V''(j\pi/2\kf)
\Bigl(1-\cos(jk\pi/2\kf)\Bigr)
\end{equation}
of harmonic excitations. The primes denote derivatives w.r.t.\
the argument. Both, $\:E_0^{\rm classical}\:$ and $\:E_0^{\rm
wc}\:$ provide rigorous lower bounds to the true ground state
energy since the quartic term of the Coulomb interaction is
positive when expanded in a power series and the Fermionic
antisymmetry constraint, ignored by Eq.~(\ref{wigqufl}),
increases the true Fermionic energy further. This latter
observation remains true also for spin carrying electrons since
spin cannot provide complete antisymmetry for symmetric spatial
wave functions for more than two particles.

Figures~\ref{e0} also include the lowest order perturbation
theory estimate
\begin{equation}\label{epert}
E_0^{\rm pert}=\frac{\vf\kf^2}{3\pi}+\frac{2\vf\kf^2}{\pi^2}\hat{V}(0)-
\frac{\vf}{2\pi^2}\int_0^{2\kf}{\rm d}k\;(2\kf-k)\hat{V}(\frac{k}{\kf})\;,
\end{equation}
obtained by taking the Hamiltonian's expectation value in the
non-interacting electron state, to obtain the positive Hartree,
and the negative exchange contribution. The variational
principle ensures that Eq.~(\ref{epert}) is a rigorous upper
bound to the ground state energy. The true ground state energy
must lie between these two bounds.

The energy densities are plotted in dimensionless units:
\begin{equation}\label{e0dim}
e_0(\kf):=\frac{E_0/L}{\kf^3/m}\quad,
\end{equation}
which has the value $\:1/3\pi\:$ without interactions. The
HF--energies $e_0^{\rm HF}$, seen in Figure~\ref{e0}~a
for $\:R/d=5.66\:$ and in Figure~\ref{e0}~b for $\:R/d=35.36\:$,
approach this value in the weakly interacting high density
limit, $\:\kf\to\infty\:$.

For densities above $\:\kf d\gtrsim 0.5\:$, corresponding to
$\:r_{\rm s}\lesssim 0.8\:$, $\:e_0^{\rm HF}\:$ agrees quite
well with the perturbative estimate (\ref{epert}). This is
despite the fact that the self consistent charge density
modulation already shows significant amplitude in this regime as
seen in Figure~\ref{cdwampl} below.

Below $\:\kf d\lesssim 0.3\:$ ($\:r_{\rm s}\gtrsim 1.3\:$),
the three approximations start to spread apart significantly.
As $\:\kf\to 0\:$ perturbation theory result diverges, $\:e_0^{\rm
pert}\sim 1/\kf\:$, while $\:e_0^{\rm wc}\sim\kf^{1/2}\:$ goes
to zero. The mean field result $\:e_0^{\rm HF}\:$ approaches a
finite value, a result which seems plausible since the
zero point quantum fluctuation energy exceeds the classical
interaction energy of the Wigner crystal at low densities
for interactions decaying faster than $\:\sim 1/|x-x'|^2\:$ at
large particle separations. We speculate that realistic quantum
wires, which never have strictly infinite range interactions,
always cross over into the hard sphere gases at sufficiently low
densities. For this latter system it is known that the ground
state energy approaches $\:e_0\to 4/3\pi\approx 0.4244\:$ (cf.\
[\onlinecite{haldane,mattis93}]), irrespective of the particle
type, Fermionic or Bosonic, and irrespective of the particle's
spin. The `radius' of the hard spheres is unimportant
when $\:\kf\to 0\:$. Among the approximations discussed above
$\:e_0^{\rm HF}\:$ is the only one that stays finite in this limit,
though the limit it approaches is larger than $\:4/3\pi\:$.

\section{TL parameter $K_{\rho}$}\label{skrho}
Figure~\ref{krho} shows
\begin{equation}\label{ke0}
1/K_{\rho}=\Kr{\frac{\pi}{2}[\kf^2e_0''(\kf)+
6(\kf e_0'(\kf)+e_0(\kf))]}^{1/2}
\end{equation}
versus $\:\kf d\:$ for $\:R/d=5.66\:$ (Figure~\ref{krho}~a) and
for $\:R/d=35.36\:$ (Figure~\ref{krho}~b). Eq.~(\ref{ke0})
follows from relation (\ref{compress}) together with
(\ref{e0dim}); the primes again denote derivatives w.r.t.\ the
arguments. Also included in these figures is the result from
expression (\ref{pop}) and the perturbative estimate
\begin{equation}\label{pert}
1/K_{\rho}^{\rm pert}=(1+(2\hat{V}(k=0)-\hat{V}(k=2\kf))/\pi)^{1/2}\quad,
\end{equation}
which follows from Eqs. (\ref{compress}) and (\ref{epert})).
Note that only this form, with the Fock term included, satisfies
the physical requirement that spinless Fermions cannot feel
contact interactions and that therefore $\:K_{\rho}\:$ equals
unity for this model. Since the factor of two in (\ref{pert})
is absent in the spinless case, Eq.\ (\ref{pert}) indeed
fulfills this Pauli principle requirement, unlike Eq.\
(\ref{pop}). Figure~\ref{krho} also includes $\:1/K_{\rho}^{\rm
cl}\:$ and $\:1/K_{\rho}^{\rm wc}\:$, calculated from the
corresponding harmonic crystal energy estimates of Eqs.\
(\ref{classic}) and (\ref{wigqufl}) using Eq.\ (\ref{compress}).

Over a wide range of densities, including the typical
experimental regime, all of these approximations coarsely agree,
though none of them can provide a rigorous bound on the exact
compressibility or $\:K_{\rho}\:$. As for the ground state
energies, the approximations start to deviate severely from one
another at smaller densities, corresponding to $\:r_{\rm
s}\gtrsim 1.5\:$. Both HF and harmonic estimates show
non-monotonic behavior of $\:1/K_{\rho}\:$ as a function of
density, in agreement with recent quantum Monte Carlo\cite{qmc}
calculations. If, as we have conjectured, $\:e_0\:$ approaches
a constant for $\:\kf\to 0\:$,
\begin{equation}\label{limit}
K_{\rho}(\kf\to 0)=(3\pi e_0(\kf\to 0))^{-1/2}\quad.
\end{equation}
Note that the HF compressibility approaches a constant in the
low density limit. Conjecturing again that at mean particle
separations exceeding the interaction range, $\:\kf\ll\pi/2R\:$,
the system crosses over into the hard core Bose gas with
$\:e_0\to 4/3\pi\:$ Eq.\ (\ref{limit}) would yield
\begin{equation}\label{klim}
K_{\rho}\longrightarrow 1/2\quad.
\end{equation}
We note, for example, that the Hubbard model approaches
(\ref{klim}) at small fillings, independent of the interaction
strength $U$. The same holds true for the Fermi gas with
contact repulsion \cite{haldane,schlottmann}. It is an
important observation that the limiting value $\:K_{\rho}^{\rm
HF}(\kf\to 0)\approx 0.29$ to $ 0.35\:$ for $\:R/d=50\ldots 8\:$
clearly {\em exceeds} 1/8 which would be the limiting value for
the extended Hubbard model \cite{schulz91} that has both on-site
and near-neighbor interactions. On the other hand, as seen in
Figure~\ref{krho}, the minimum value for $\:K_{\rho}\:$ at about
$\:\kf R\sim 1\:$ is considerably smaller than 1/2, so that,
contrary to the Hubbard model, the limit (\ref{klim}) would have
to be approached {\em from below\/} with decreasing carrier
densities in quantum wires.

Upon inspecting Figures~\ref{krho} more closely a regime
can be identified at densities somewhat above
the maximum of $\:1/K_{\rho}^{\rm HF}\:$ where
$\:1/K_{\rho}^{\rm HF}\:$ {\em exceeds\/} $\:1/K_{\rho}^{\rm
pert}\:$. As seen in Figure~\ref{cdwampl},
the relative increase in
stiffness appears along with the occurrence of significant
$4\kf$--periodic contribution to the charge density modulation.
Figure~\ref{cdwampl} shows the two lowest Fourier coefficients
\begin{equation}\label{densfcoeff}
\varrho_j\equiv\varrho_{\uparrow}(q=j2\kf)=
(-1)^j\varrho_{\downarrow}(q=j2\kf)=\varrho_{-j}
\end{equation}
for $j=1,2$ in units of the mean density
$\:\varrho_0=2\kf/\pi\:$. In view of (\ref{densfcoeff}), which
follows from Eq.~(\ref{antif}) of the Appendix, $4\kf$--periodic
modulations of the charge density
$\:\varrho_{\uparrow}(x)+\varrho_{\downarrow}(x)\:$ are given by
the $j=2$ contribution in Figure~\ref{cdwampl}. The appearance
of a substantial $j=2$ Fourier component, at $\:\kf d \sim
0.5\:$, marks the crossover from spin-density-wave to Wigner
crystal self-consistent solutions of the HF equations. A
similar conclusion has been drawn from the extremely slow
spatial decay of the density--density correlation function in
the presence of long range interactions \cite{schulz93} and from
recent quantum Monte Carlo studies \cite{qmc}. With smaller
$\:1/R\:$ this regime of Wigner crystal-like states marked by
enhanced stiffness extends down to smaller densities and becomes
more pronounced. The variation of $\:1/K_{\rho}^{\rm HF}\:$
with $R$ is depicted in Figure~\ref{krhor} for the density
$\:\kf d=0.15\:$. At $\:\kf R\gg 1\:$ all of the approximate
estimates are consistent with the logarithmic increase
$\:1/K_{\rho}\sim\sqrt{\log R/d}\:$ suggested by perturbation
theory. For $\:\kf R\gg 1\:$, the electrostatic energy is so
dominant that the energy and $K_{\rho}$ are relatively
insensitive to correlations.

\section{Spin Sector}\label{spinsector}
As mentioned already in the introduction, it is much more
difficult to estimate how interactions influence the
spin sector, and particularly its low energy TL parameter
$v_{\sigma}$, than it is to estimate the charge sector parameter.
In the model originally proposed by Luttinger
\cite{luttinger} with left and right going particles treated as
distinguishable, the spin velocity is unrenormalized
\cite{overhauser}, $v_{\sigma}=\vf$, because the exchange term
vanishes leaving magnetic properties of the system independent
of interactions. For the Hubbard model, on the
other hand, is known that the spin velocity
\cite{coll} is dependent on interactions and particle density,
$\:n\:$, vanishing like $ n^2\:$ at small density
$\:n\:$ for any finite interaction strength. The spin TL parameter
is related to a thermodynamic quantity, the magnetic susceptibility, by
\cite{frahm90}
\begin{equation}\label{chivs}
\chi\equiv 4(\pi^2\vf\partial_m^2e_0(m))^{-1}=
\frac{2K_{\sigma}}{\pi v_{\sigma}}
\end{equation}
where $\:m=(n_{\uparrow}-n_{\downarrow})/n\:$ is the
magnetization per particle and $e_0$ is the dimensionless ground
state energy density as defined in Eq.~(\ref{e0dim}). Relation
(\ref{chivs}) actually holds for any interacting electron system
in the single channel TL phase \cite{schulz93a,voit}.
Evaluating $e_0(m)$ for the microscopic model perturbatively
would give
\begin{equation}\label{vswrong}
\tilde{v}_{\sigma}/\vf=1-\hat{V}(2\kf)/\pi\;,
\end{equation}
using (\ref{chivs}) and $K_{\sigma}=1$. Alternatively one also
could impose a spin current $\:\ell=(n_{\mbox{\tiny R}\uparrow}-
n_{\mbox{\tiny R}\downarrow}-n_{\mbox{\tiny
L}\uparrow}+n_{\mbox{\tiny L}\downarrow})/n\:$ per particle
($n_{\mbox{\tiny R/L}s}$ are right/left moving densities of
spin $s$) and measure the change in ground state energy
\begin{equation}\label{spindrag}
\chi_{\ell}\equiv 4(\pi^2\vf\partial_{\ell}^2e_0(\ell))^{-1}
=\frac{2}{\pi v_{\sigma}K_{\sigma}}\;.
\end{equation}
Perturbatively this gives an unchanged spin velocity, a result
that simply reflects the fact that lowest order perturbation
theory cannot describe drag effects \cite{zheng} between the
density fluctuations of opposite spins and thus leaves the
system Galilean invariant in spin sector.
Solving equations (\ref{chivs}) and (\ref{spindrag}) for
$K_{\sigma}$ yields $K_{\sigma}>1$ as a perturbative result for
repulsive interactions which would contradict SU(2) invariance in a
TL model. To enforce the SU(2) symmetry we can combine equations
(\ref{chivs}) and (\ref{spindrag}) and solve for $v_{\sigma}$ by
eliminating $K_{\sigma}$. The result,
\begin{equation}\label{vsright}
v_{\sigma}^{\rm pert}=\frac{2}{\pi}(\chi\chi_{\ell})^{-1/2}=
\vf\sqrt{1-\hat{V}(2\kf)/\pi},
\end{equation}
indeed agrees clearly better with the QMC data \cite{qmc} than
(\ref{vswrong}). Eq.~(\ref{vsright}) is included in
Figure~\ref{vsvalues}. At $\:\hat{V}(2\kf)=\pi\:$ the
perturbative estimate $v_{\sigma}^{\rm pert}$ vanishes and
for smaller $\kf$ the Fock term in (\ref{vsright}) favors a spin
polarized ground state. This result contradicts very general arguments
that guarantee a non-magnetic ground state for any non-singular
pair interaction potential in one dimension \cite{liebmattis}.
The true spin velocity should stay
positive and approach zero only at vanishing particle density.

That the extraction of spin velocities from HF calculations
is less reliable than the extraction of charge TL parameters,
is already clear because of the incorrectly broken spin-rotational
invariance in the HF ground state In the HF spin-density wave
state we evaluate the spin-susceptibility by polarizing spins
along the quantization axes. We can consider only
cases with rational ratios of the spin-up and spin-down carrier
densities. Because the periods of the spin-up and spin-down
density waves differ in these solutions, it is more convenient
to use a real space basis, discretizing space
$\:\psi(x=x_i)\to\psi(i)\:$ as described in
Appendix~\ref{spindetails}. Self consistent solutions are shown
in Figure~\ref{pdensities}. At finite magnetization this
structure contains now `defects', reflecting the loss of the
$4\kf$--periodic component in the charge density modulations
\cite{carmelo93b}. At least $\:N=44\:$ electrons have been
considered on $\:M=401\:$ grid points, the smaller particle
densities are based on $\:N=84\:$ and $\:M=801\:$ to avoid
lattice artifacts to a high accuracy. These sizes are clearly
beyond what presently can be treated with numerical many body
approaches, like quantum Monte Carlo, but pose no problem here.
Spin velocities obtained from $E_0^{\rm HF}(m)/L$ by virtue of
Eq.~(\ref{chivs}) are included in Figure~\ref{vsvalues}. Below
$\kf d=0.2$ it is very difficult to extract positive spin
velocities. We see that selfconsistency pushes the point of
vanishing spin velocity and the (erroneous) transition into a
ferromagnetic ground state down to smaller densities compared to
the perturbative estimate in Eq. (\ref{vsright}), but the
transition still occurs.

An alternative attempt to estimate the spin velocity starts from
the argument that the electron spin sector would evolve at low
particle densities towards that of an antiferromagnetic
Heisenberg spin chain, as suggested by the staggered spin
density profile found in the mean field solution,
Fig~\ref{densities}. This argument is also suggested by the
pronounced antiferromagnetic correlations found in the TL liquid
spin sector, particularly for long range interactions
\cite{schulz93}, and in finite pieces of one-dimensional wires
\cite{jjwh}. Antiferromagnetic spin chains are known to
represent microscopic models, such as the Hubbard model, at low
energies and have been intensively investigated for instance by
employing manifestly SU(2) spin rotation invariant non--Abelian
bosonization \cite{agsz,nonabelian}.

In the antiferromagnetic Heisenberg chain spin excitations
(magnons) move at velocity
\[
v_{\sigma}=\pi^2J/4\kf
\]
where $\:J\:$ is the nearest neighbor coupling constant. One
possibility to guess the magnitude of $\:J\:$ is to compare the
HF estimates for the ground state energy $\:e_0^{\rm HF}\:$ of
unpolarized electrons with the ground state energy $\:e_0^{\rm
pol}\:$ of fully spin polarized electrons. For the
antiferromagnetic Heisenberg chain this energy difference,
$\:J(1+\ln 2)\:$ per spin, \cite{bethe} is known exactly.
Equating the energy differences gives
\begin{equation}\label{jdet}
J=\frac{\pi}{1+\ln 2}\frac{\vf\kf}{2}(e_0^{\rm pol}-e_0^{\rm HF})
\end{equation}
from which
\begin{equation}\label{vspol}
v_{\sigma}^{\rm J}/\vf=\pi^3(e_0^{\rm pol}-e_0^{\rm HF})/8(1+\ln 2)
\end{equation}
follows. Eq.~(\ref{vspol}) is included in
Figure~\ref{vsvalues}. The transition into the spin polarized
ground state occurs at $\kf d=0.19$ ($\:r_{\rm s}=2.07\:$) for
$R/d=5.66$. Equation (\ref{jdet}) can be checked for
consistency in the non-interacting limit, $\:\kf\to\infty\:$,
where $\:v_{\sigma}\to\vf\:$. Magnons would move at velocity
$\:\vf\:$ if $\:J=4\vf\kf/\pi^2=0.41\:\vf\kf\:$. On the other
hand, $\:(e_0^{\rm pol}-e_0^{\rm HF})\to 1/\pi\:$ in this limit
so that (\ref{jdet}) yields $\:J=0.30\:\vf\kf\:$. In view of
the fact that the weak interaction limit is poorly described by
the antiferromagnetic spin chain this picture seems amazingly
consistent.

Recently, Calmels and Gold have calculated magnetic
susceptibilities of quantum wires \cite{calmels}, though for a
different microscopic interaction, using standard heuristic
approximations from electron gas theory \cite{singwi}. By
virtue of Eq.~(\ref{chivs}) these data allow us to extract spin
velocities $\:v_{\sigma}^{\rm CG}\:$ which turn out to be
slightly larger than our $\:v_{\sigma}^{\rm HF}\:$ values. Note
that the perturbative estimate shown in Figure~2 of Ref.\
[\onlinecite{calmels}] uses Eq.~(\ref{vswrong}) while in
Figure~\ref{vsvalues} Eq.~(\ref{vsright}) is included. Compared
with (\ref{vsright}) the data $\:v_{\sigma}^{\rm CG}\:$ are
smaller than $\:v_{\sigma}^{\rm pert}\:$ at large densities,
like the Hartree-Fock data. This latter property does not agree
with the behavior obtained using QMC \cite{qmc} where
$\:v_{\sigma}^{\rm QMC}\:$ instead exceeds $\:v_{\sigma}^{\rm
pert}\:$. We conclude that HF and other approximations of the
mean field type can provide only a qualitative guideline to
$\:v_{\sigma}\:$. All of these attempts, however, agree in
predicting spin velocities that depend on the interaction and
{\it decrease} with increasing interaction strength. This
result calls attention to the frequent assumptions in the
literature that interactions not explicitly depending on spin
would leave $\:v_{\sigma}\:$ unchanged \cite{mahan}. This
result should show up in current experiments, such as those
described in Ref.~[\onlinecite{yacoby}], where typical values
for $\:\kf d\approx 0.3\:$ are in the regime investigated here.
As already pointed out in the introduction, this parameter
should influence measurable quantities, such as the
spin-splitting enhancement factor, Raman scattering in
depolarized configuration \cite{goni}, spin transport properties
\cite{balents}, and Rashba precession\cite{whrashba}.

Let us now discuss the low density limit using our conjecture
that quantum wires become equivalent to the on-site Hubbard model in the
limit of small particle densities. Hubbard model (lattice
constant $\:a\:$) parameters, $t=\vf/2\kf a^2$ and
$U=\hat{V}(k=0)\vf/a$, can be related to microscopic
parameters by equating the effective mass and the Coulomb
barrier for a two electron exchange, $\hat{V}(k)$ is defined in
Eq.(\ref{vfour}). To leading order in $t/U$ the spin velocity of
the Hubbard model $\:v_{\sigma}^{\rm HM}\:$ is \cite{coll}
\[
v_{\sigma}^{\rm HM}\quad
\mbox{\raisebox{1ex}{$\mbox{\tiny$U\to\infty$}\atop\longrightarrow$}}
\quad\frac{2\pi at^2}{U}\Kr{1-\frac{\sin 4\kf a}{4\kf a}}
\]
so that
\begin{equation}\label{hcvs}
\frac{v_{\sigma}^{\rm HM}}{\vf}=4\pi/3\hat{V}(k=0)
\end{equation}
for small $\:\kf a\:$, where the lattice constant is irrelevant.
Note that $\:\hat{V}\sim\vf^{-1}\:$ and thus $\:v_{\sigma}^{\rm
HM}\propto\kf^2\:$. This result agrees with the strong
interaction limit of the continuum version, the electron gas
with repulsive contact interactions \cite{schlottmann}.
Restoring quantum wire parameters Eq.~(\ref{hcvs}) translates
into
\begin{equation}\label{vslim}
\frac{v_{\sigma}}{\vf}\quad
\mbox{\raisebox{1ex}{$\mbox{\tiny$\kf\to 0$}\atop\longrightarrow$}}
\quad\frac{2\pi}{3}\frac{\kf a_{\rm B}}{\ln(2R/d)}
\end{equation}
for $\:R/d\gg 1\:$. The available QMC data are consistent with
(\ref{vslim}), though, as in the charge sector, they are
not conclusive enough to really confirm the low density
equivalence.

\section{Summary and Discussion}
Many non-trivial theoretical predictions exist in the literature
on various measurable low energy properties of single channel
quantum wires that are based on the TL model which describes the
non-Fermi liquid behavior of these systems. Much in the spirit
of the Landau theory of Fermi liquids all these predictions
depend only on few phenomenological parameters. In the case of
quantum wires with only one subband occupied there is only one
parameter per degree of freedom, $\:K_{\rho}\:$ for the charge
sector and $\:v_{\sigma}\:$ for the spin sector. In the absence
of interactions these parameters assume the values
$\:K_{\rho}=1\:$ and $\:v_{\sigma}=\vf\:$. In this work we have
investigated how the microscopic pair potential $\:V(x-x')\:$
changes $\:K_{\rho}\:$ and $\:v_{\sigma}\:$. We have considered
a realistic, tractable, and sufficiently general form for
$\:V(x-x')\:$, Eq.~(\ref{xinteraction}), that depends on the
diameter $\:d\:$ of the quantum wire which is measurable through
the subband energy, and the distance to the nearest metallic
gates $\:R\:$ as given by the sample lay out. Our approach is
to relate the two TL parameters to thermodynamic quantities
which we estimate on the basis of self consistent, unrestricted
Hartree-Fock (HF) approximations for the ground state energy.

In the charge sector this strategy is found to yield reasonably
accurate results. At densities corresponding to $\:r_{\rm
s}\lesssim 1.3\:$ we confirm applicability of the perturbative
formula (\ref{pert}) for $\:K_{\rho}\:$. This regime includes
most of the experiments based on semiconducting
heterostructures \cite{tarucha,goni}. At somewhat smaller
densities Eq.~(\ref{pert}) even {\em over\/}estimates
$\:K_{\rho}\:$. In this regime we find enhanced stiffness
compared to perturbation theory, signaling the close proximity
of a Wigner--crystal state. For this reason quantitative
corrections to Eq.~(\ref{pert}) may arise when
$\:R>\!\!\!>\!\!\!>d\:$, for example in carbon nanotubes.
In quantum wires fabricated on the basis of semiconducting
hetero--structures with gates, the perturbative formula may be
used even down to densities $\:2\kf/\pi\approx 1/R\:$. The
proximity to the Wigner--crystal state competes with the finite
interaction range in these systems. Irrelevant or marginal
operators in the microscopic Hamiltonian, such as non-linear
single particle dispersion or backward scattering in the spin
sector turn out to be unexpectedly inefficient to renormalize
the TL parameters in the charge sector up to moderate interaction
strengths. With decreasing density the values for $\:K_{\rho}\:$
clearly fall short of 1/8 which is the minimum assumed by the
extended Hubbard model including repulsions on neighboring lattice
sites and often is considered to emulate models of finite
interaction range.

At smaller densities, however, does $\:K_{\rho}^{\rm pert}\:$
become poor. In particular, it does not reproduce the
non-monotonic behavior of $\:K_{\rho}\:$ found in the HF
approximation with a minimum as a function of density.
Eventually, as $\:\kf\to 0\:$ we conjecture that quantum wires
approach the universality class of the Hubbard model with only
on-site repulsion and that $\:K_{\rho}\to 1/2\:$ in that limit,
though, unlike the Hubbard model, this limiting value should be
approached {\em from below\/} as the particle density is
lowered.

The non-monotonic dependence of $\:K_{\rho}\:$ on density
predicted here should show up in any of the power laws
\cite{kanefisher} revealed by pseudo gaps in the density of
states $\:\nu\:$. Examples include the current for tunneling
into the end ($\:\nu(\omega)\sim\omega^{(1/K_{\rho}-1)/2}\:$) or
into the middle ($\:\nu(\omega)\sim\omega^{(K_{\rho}+1/K_{\rho}-
2)/4}\:$) of a single mode wire (assuming $\:K_{\sigma}=1\:$),
and the current $\:I(V)\sim V^{1/K_{\rho}}\:$ flowing through a
single tunnel barrier along the wire at small voltages $\:V\:$.
Experimental observation of this non-monotonic dependence of the
exponent would give direct experimental access to the
microscopic range of the electron--electron interaction; the
position and the height of the maximum in $\:K_{\rho}^{-1}\:$
both depend on $\:R\:$.

Our approach is less successful in estimating the spin sector TL
parameter $\:v_{\sigma}\:$, at least when it differs
considerably from $\vf$. We have discussed perturbation theory
and tried to obtain meaningful estimates for $\:v_{\sigma}\:$
from the HF spin density wave states. The resemblance to the
antiferromagnetic Heisenberg spin chain, suggested by
correlation function considerations, suggests that exchange
coupling strengths, and therefore spin velocities, can be
estimated by comparing the ground state energies of unpolarized
and fully spin polarized electrons. None of these variants lead
to results of the same quantitative reliability as those
obtained from $K_{\rho}^{\rm HF}$. Conjecturing again a cross
over into the universality class of the Hubbard model in the
limit of $\:\kf\to 0\:$ yields the prediction of a linear
dependence of the {\em relative\/} spin velocity
$\:v_{\sigma}/\vf\propto\kf/V_0\:$ on the particle density.
$\:V_0\:$ is the zeroth Fourier component of the interaction
$\:V(x-x')\:$.

It is important to know the spin velocities for attempts to
realize `spintronic' devices where spins rather than charges are
transported \cite{spintransistor}, using for example the Rashba
spin precession mechanism \cite{dattadas} through quasi
one-dimensional constrictions \cite{tsukagoshi99}. Here, in
agreement with QMC estimates \cite{qmc}, we have collected
strong evidence that spin density excitations move at speeds
considerably slower than the Fermi velocity, already in present
day devices \cite{yacoby}, where $\:\kf d\approx 0.3\:$.

\vspace{1cm}

\noindent{\bf Acknowledgement}\\
We would like to thank Ulrich Z"ulicke, Charles Creffield,
and Hermann Grabert for valuable discussions.
WH acknowledges the enjoyable hospitality of
Indiana University at Bloomington, and also kind
hospitalities of the University of Freiburg, and the King's
College London where parts of this work have been carried out.
Support has been received from the Deutsche Forschungsgemeinschaft
(HA 2108/4-1), from the British EPSRC, and from the National Science
Foundation under grant DMR0105947.

\begin{appendix}
\section{Hartree-Fock Theory}\label{hfthy}
To formulate the mean field theory we introduce single particle
wave functions $\:\psi\:$ solving the Schr"odinger equation
\begin{equation}\label{single}
\Kg{-\frac{v_{\rm F}}{2k_{\rm F}}\partial_x^2+
\sum_{s'}V_{s'}^{\rm H}(x)}\psi_{ks}(x)-
\int_0^L{\rm d}x'\;V_{s}^{\rm E}(x,x')\psi_{ks}(x')=
\ve_{ks}\psi_{ks}(x)
\end{equation}
with the Hartree
\begin{equation}\label{hartreepot}
V_{s'}^{\rm H}(x)=\frac{L}{2\pi}\int_0^L{\rm d}x'\;
V(x-x')\int_{-k_{\rm F}}^{k_{\rm F}}{\rm d}k\;|\psi_{ks'}(x')|^2
\end{equation}
and the non-local exchange
\begin{equation}\label{exchangepot}
V_{s}^{\rm E}(x,x')=\frac{L}{2\pi}V(x-x')\int_{-\kf}^{\kf}{\rm d}k\;
\psi_{ks}^*(x')\psi_{ks}^{}(x)
\end{equation}
potentials, which itself depend on $\:\psi\:$ and thus have to
be obtained selfconsistently; $\:s=\uparrow,\downarrow\:$ are
spin quantum numbers. Occupation of only the lowest subband is
assumed together with periodic boundary conditions for the wire
of length $\:L\:$. Parabolic dispersion for the kinetic energy
in (\ref{single}) is described by a band mass $\:m=\kf/\vf\:$.
The kinetic energy is not linearized.

For some of our calculations, particularly those focussing on
properties of the spin sector, cf.\ Section~\ref{spinsector}, we
solved the HF equations (\ref{single}) directly in real space,
using
\begin{equation}\label{xinteraction}
V(|x|)=\frac{e^2}{\epsilon}(\frac{1}{\sqrt{x^2+d^2}}-
\frac{1}{\sqrt{x^2+d^2+4R^2}})
\end{equation}
in (\ref{hartreepot},\ref{exchangepot}) and a lattice grid of at
least 401 points. Any of the results for the charge sector
can be obtained either using a real space basis as also,
slightly more efficiently, in $k$--space, introduced now. Expanding
\begin{equation}\label{bloch}
\psi_{ks}(x)=\ex{kx}\sum_ju_{j,k,s}\:\ex{j2\kf x}
\end{equation}
into Bloch waves, and similarly the periodic potentials
(\ref{hartreepot}) and (\ref{exchangepot}), yields HF
equations for the coefficients $\:u_{j,k,s}\:$
\begin{eqnarray}\label{schrod}
0&=&\d\Ke{\frac{1}{2}(2j-\frac{k}{k_{\rm F}})^2-
\frac{\ve_{ks}}{\kf\vf}}u_{j,k,s}+
\frac{L}{2k_{\rm F}\pi}\sum_{j'j''}u_{j'',k,s}
\int_{-k_{\rm F}}^{k_{\rm F}}{\rm d}k'\\[3ex]
&&\times\d\Kg{\hat{V}\Bigl(2(j-j'')\Bigr)\:\sum_{s'}u_{-j+j'+j'',k',s'}^*\;
u_{j',k',s'}-\hat{V}\Bigl(2(j-j')-\frac{k}{k_{\rm F}}+\frac{k'}{k_{\rm F}}
\Bigr)\:u_{-j+j'+j'',k',s}^*\;u_{j',k',s}}\quad.\nonumber
\end{eqnarray}
For each $\:s=\pm 1\:$ and $\:k=-\kf\ldots\kf\:$ inside the
Brillouin zone this is an eigenvalue equation for matrices
indexed by the band indices $\:j\:$ which, however, inside the
curly bracket, depends on the solution of (\ref{schrod}).
Within the `unrestricted' HF scheme we allow for charge and spin
density wave solutions breaking the symmetry of continuous
translations and thereby lower the ground state energy. Solutions
are found to show $\:4\kf$--periodic oscillations of the charge
density $\:\varrho(x)=\varrho_{\uparrow}(x)+\varrho_{\downarrow}(x)\:$
where
\begin{equation}\label{antif}
\varrho_{\uparrow}(x)\equiv\int_{-k_{\rm F}}^{k_{\rm F}}{\rm d}k\;
|\psi_{k\uparrow}(x)|^2=\varrho_{\downarrow}(x+\frac{2\pi}{4k_{\rm F}})
\quad.
\end{equation}
We solved Eq.~(\ref{schrod}) iteratively, starting
with a sinusoidal spin density wave
$\:u_{j,k,s}^{(0)}=\delta_{j,0}/\sqrt{2}+s\delta_{|j|,1}/2\:$.
The final solution always obeys $\:u_{j,k,\uparrow}=(-1)^j
u_{j,k,\downarrow}\:$ which in view of (\ref{antif}) yields
$\:2\kf$--periodic modulations of the spin density
$\:\varrho_{\uparrow}(x)-\varrho_{\downarrow}(x)\:$. A typical
density modulation at stronger interaction is shown in
Figure~\ref{densities}.

The single particle energies $\:\ve_{ks}\:$, obtained with
(\ref{schrod}), determine the ground state energy
\begin{eqnarray}\label{e0dens}
\d\frac{E_0}{L}=\sum_s\int_{-\kf}^{\kf}\frac{{\rm d}k}{2\pi}\;
\ve_{ks}&-&\d\frac{1}{2}\sum_{ss'}\int_{-\kf}^{\kf}\frac{{\rm d}k}{2\pi}
\int_0^L{\rm d}x\;V_{s'}^{\rm H}(x)|\psi_{ks}(x)|^2\\[3ex] \nonumber
&+&\d\sum_s\int_{-\kf}^{\kf}\frac{{\rm d}k}{2\pi}\int{\rm d}x\int_0^L
{\rm d}x'\;V_{s}^{\rm E}(x,x')\psi_{ks}^*(x)\psi_{ks}^{}(x')\quad.
\end{eqnarray}
Half of the interaction has to be subtracted to repair for its
double counting in (\ref{schrod}). Differentiating $\:E_0/L\:$
twice w.r.t.\ $\:\kf\:$ yields our estimate for $\:K_{\rho}\:$,
according to (\ref{compress}).

Most of the results are obtained for 82 $\:k$--points in the
Brillouin zone (in some cases of very small densities we increased
this number to 234). The Milne--rule, being accurate to 7--th
order in the spacing between $k$--points, is used for the
$k$--integrations. We included $\:j=-3,\ldots,3\:$ bands though
in most cases $\:j=-2,\ldots,2\:$ would have sufficed due to the
rapid decay of the Coulomb interaction in $k$--space.

\section{Hartree-Fock for susceptibilities}\label{spindetails}
In spin space we use a lattice representation
$\:\psi(x=x_i)\to\psi(i)\:$ of the Hamiltonian. The $M\times M$
matrices $H_{ij}$, representing Eq.~(\ref{single}), contain
contributions from the kinetic energy $\:H_{ii}=2(M/\pi
N)^2\vf\kf\:$ and $\:H_{ii\pm 1}=-(M/\pi N)^2\vf\kf\:$, the
(local) Hartree-term $\:H_{ii}=\sum_{kjs}\tilde{V}(|i-j|L/M)
|\psi_{ks}(j)|^2\:$, and the (non-local) Fock-term
$\:H_{ij}=-\sum_{k}\tilde{V}(|i-j|L/M)\psi_{ks}(i)\psi_{ks}(j)\:$,
the latter acting only on spin-$s$ wave functions. Here,
$\:\tilde{V}(x)\equiv \sum_{l=-\infty}^{\infty}V(x+lL)\:$
accounts for periodic boundary conditions ($V(x)$ is defined in
(\ref{xinteraction})) and the real and normalized eigenvectors
$\psi_{ks}(j)$ of $H_{ij}$ are indexed by their spin
$s=\pm 1$ and momentum $-k_{{\rm F}s}\le k\le
k_{{\rm F}s}=(1+sm)\kf$ with $k$ being an integer multiple of
$2\pi/L$ and $m$ the magnetization per particle. The Hartree-Fock
approximation to the ground state energy $E_0$ then is obtained
as in Eq.~(\ref{e0dens}).
\end{appendix}

\begin{raggedright}

\end{raggedright}
\renewcommand{\baselinestretch}{1}\large\normalsize

\newpage\noindent{\bf Figures}

\begin{figure}[h]\caption{\label{e0}}
\fcap{Figure~3: Ground state energy densities
$\:e_0(\kf)=(E_0/L)/(\kf^3/m)\:$ for $\:R/d=5.66\:$ (a)
and for $\:R/d=35.36\:$ (b) in HF (solid). The perturbative
estimate (Eq.~(\ref{epert}), long-dashed) establishes an upper
bound while the harmonic chain estimates, both, omitting
(Eq.~(\ref{classic}), dotted) or including quantum fluctuations
(Eq.~(\ref{wigqufl}), dashed) both establish lower bounds to the
true ground state energy.}
\end{figure}

\begin{figure}[h]\caption{\label{krho}}
\fcap{Figure~4: $\:1/K_{\rho}\:$ versus $\:\kf d\:$
for $\:R/d=5.66\:$ (a) and for $\:R/d=35.36\:$ (b).
The same approximations are included as in Figure~\ref{e0},
together with the commonly used formula (\ref{pop}),
dash--dotted.}
\end{figure}

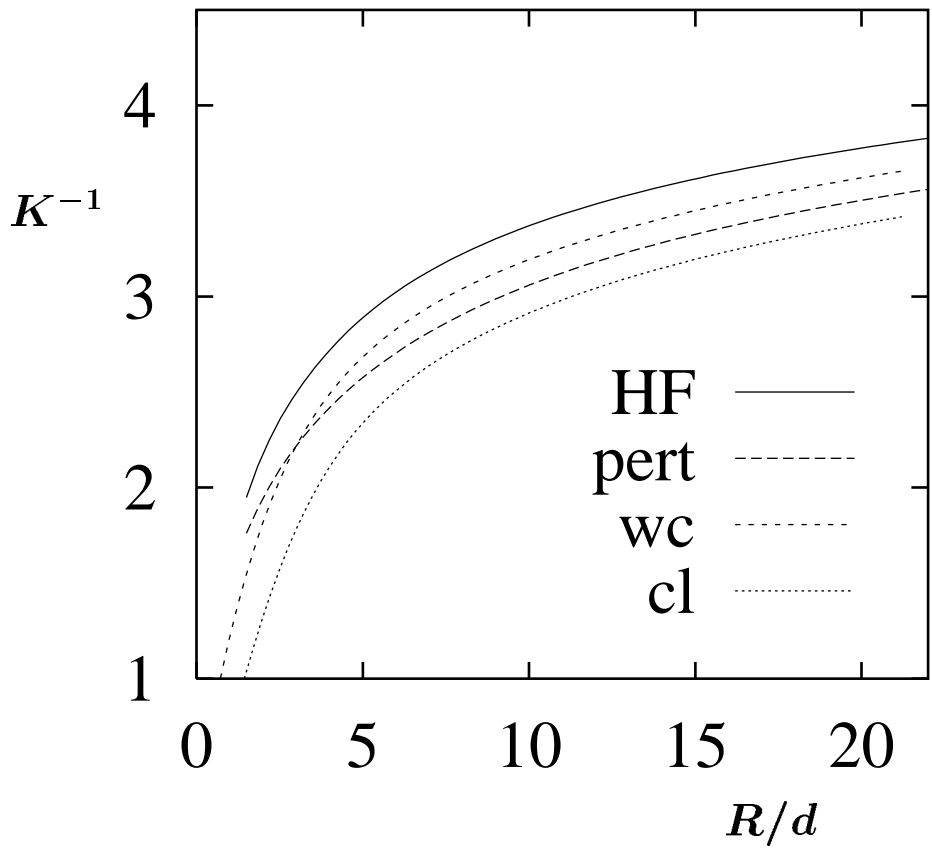
\begin{figure}[h]\caption{\label{krhor}}
\fcap{Figure~5: $\:1/K_{\rho}\:$ versus $\:R/d\:$
for $\:\kf d=0.15\:$ in HF (solid), perturbation theory
(Eq.~(\ref{pert}), long-dashed), and for the harmonic chain
(Eq.~(\ref{classic}), dashed) and (Eq.~(\ref{wigqufl}), dotted).}
\end{figure}

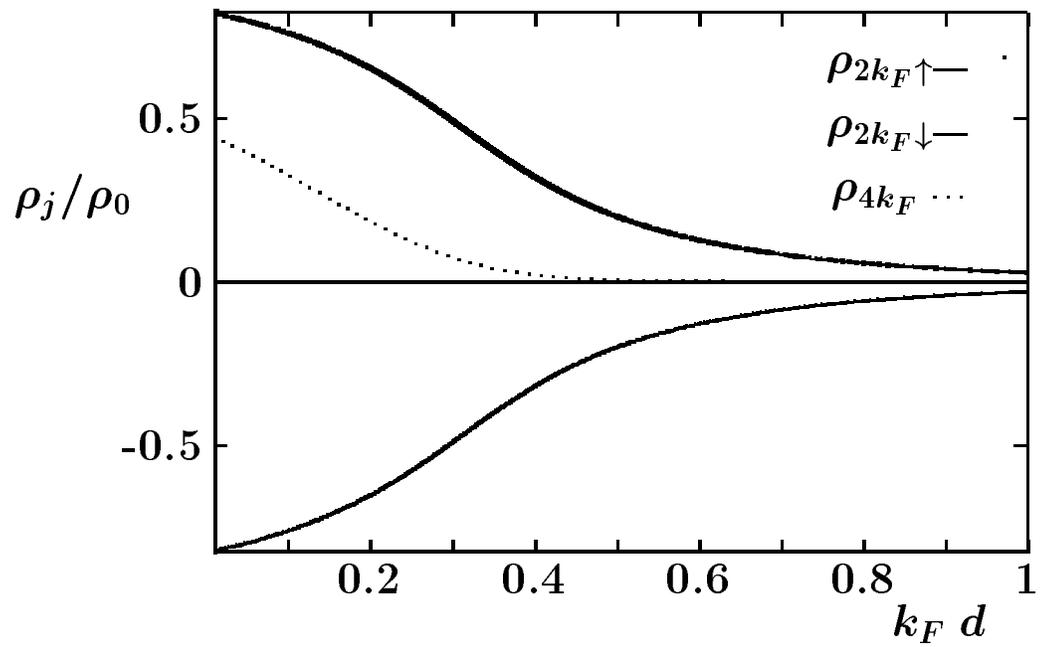
\begin{figure}[h]\caption{\label{cdwampl}}
\fcap{Figure~6: Amplitudes for the $2\kf$ and $4\kf$--periodic
components of the charge density modulations $\rho_1$ and
$\rho_2$ versus $\:\kf d\:$ for $R/d=35.36$ in units of
$\rho_0=2\kf/\pi$.}
\end{figure}

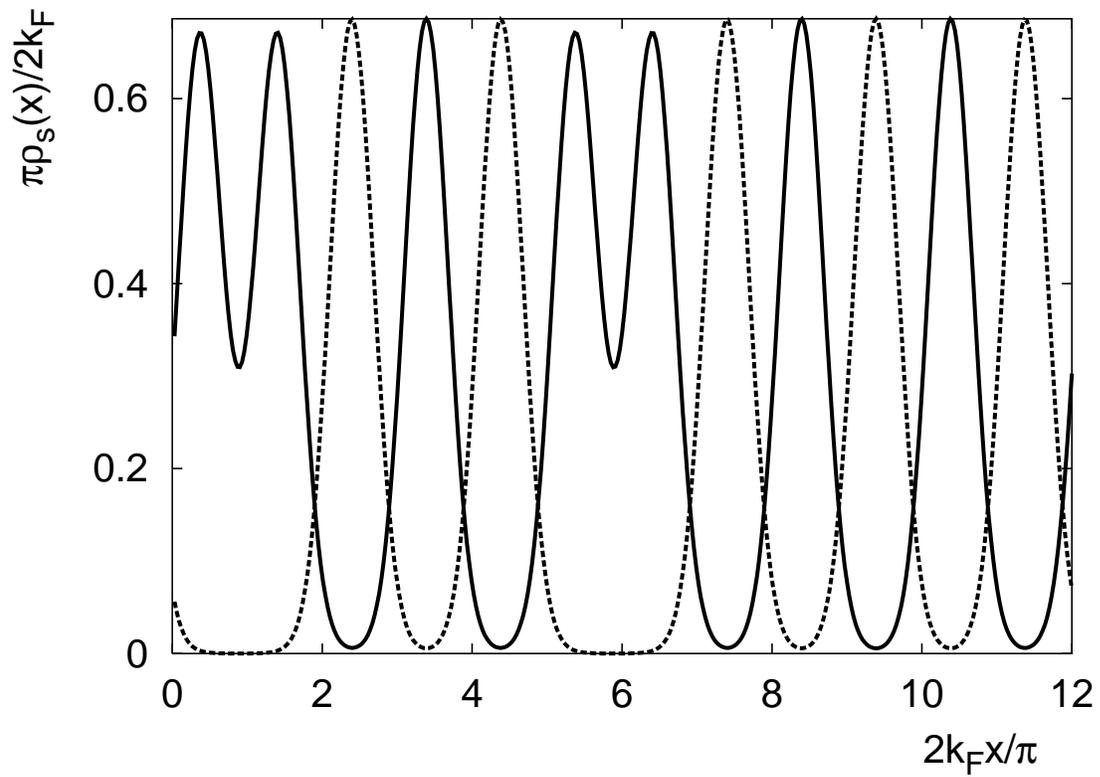
\begin{figure}[h]\caption{\label{pdensities}}
\fcap{Figure~7: Charge densities $\:n_{\uparrow}(x)\:$ (solid)
and $\:n_{\downarrow}(x)\:$ (dashed) along the wire $\:x\:$ in
units of the mean electron spacing, as in Figure~\ref{densities}
but for finite magnetization $m=1/6$ per spin.}
\end{figure}

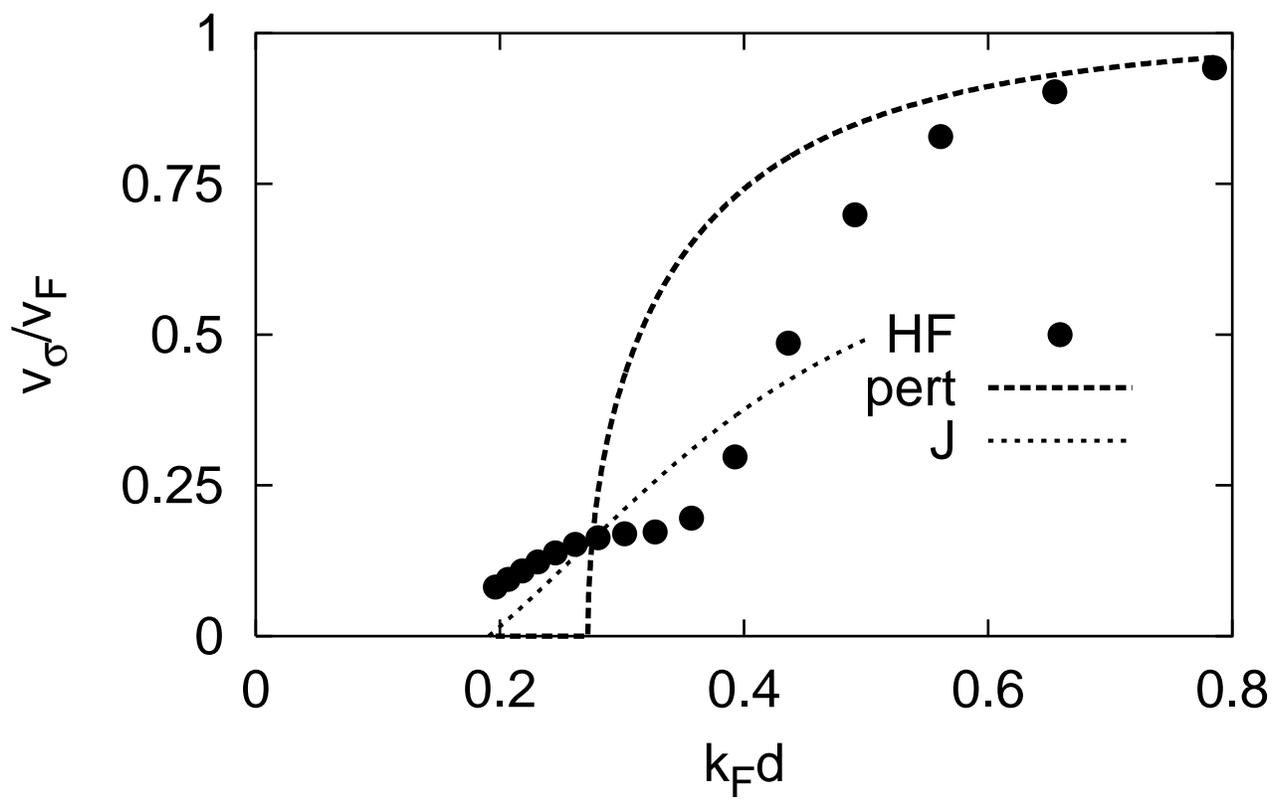
\begin{figure}[h]\caption{\label{vsvalues}}
\fcap{Figure~8: Estimates to spin velocities
$\:v_{\sigma}/\vf\:$, based on the selfconsistent
Hartree-Fock solution~: {\sf `HF'},
on perturbation theory, Eq.~(\ref{vsright})~: {\sf `pert'}, and
on the comparison with the antiferromagnetic Heisenberg model,
Eq.~(\ref{vspol})~: {\sf `J'}.}
\end{figure}

\newpage\vspace*{\fill}\hspace*{-2cm}\includegraphics{fig1}

FIGURE 1 %\hspace{1cm}H\"AUSLER

\newpage\vspace*{\fill}\hspace*{-2cm}\includegraphics{fig2}

FIGURE 2 %\hspace{1cm}H\"AUSLER

\newpage\vspace*{\fill}\hspace*{-2cm}\includegraphics{fig3a}

FIGURE 3a %\hspace{1cm}H\"AUSLER

\newpage\vspace*{\fill}\hspace*{-2cm}\includegraphics{fig3b}

FIGURE 3b %\hspace{1cm}H\"AUSLER

\newpage\vspace*{\fill}\hspace*{-2cm}\includegraphics{fig4a}

FIGURE 4a %\hspace{1cm}H\"AUSLER

\newpage\vspace*{\fill}\hspace*{-2cm}\includegraphics{fig4b}

FIGURE 4b %\hspace{1cm}H\"AUSLER

\newpage\vspace*{\fill}\hspace*{-2cm}\includegraphics{fig5}

FIGURE 5 %\hspace{1cm}H\"AUSLER

\newpage\vspace*{\fill}\hspace*{-2cm}\includegraphics{fig6}

FIGURE 6 %\hspace{1cm}H\"AUSLER

\newpage\vspace*{\fill}\hspace*{-2cm}\includegraphics{fig7}

FIGURE 7 %\hspace{1cm}H\"AUSLER

\newpage\vspace*{\fill}\hspace*{-2cm}\includegraphics{fig8}

FIGURE 8 %\hspace{1cm}H\"AUSLER


\begin{thebibliography}{99}
\bibitem{circcurr} Galilei invariance ensures that no physical
 property is affected by $\:\nu(k)\to\nu(k+q)\:$ for any $q$;
 $\:\nu(k)\:$ are density modulations of wave vector $k$ and
 $\:\nu=\rho,\sigma\:$. For the TL follows that $K_{\rho}v_{\rho}$
 does not change with interactions depending on the distance.
\bibitem{pertmean} Throughout this work by `perturbation theory' we mean
 the expressions obtained in lowest order.
\end{thebibliography}
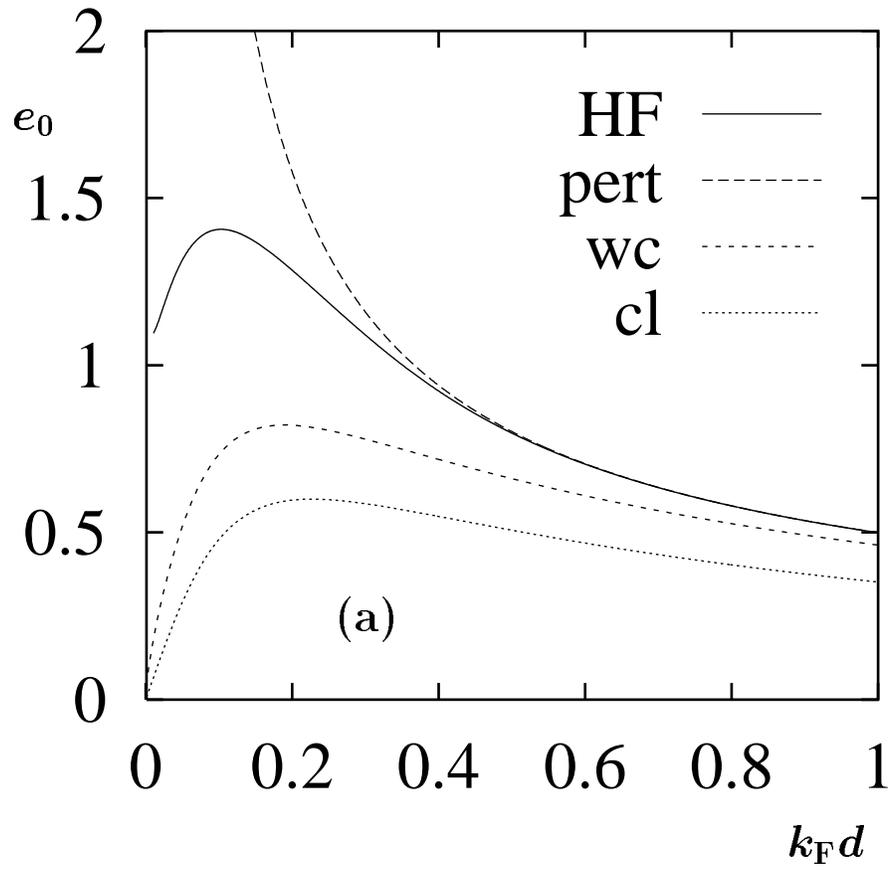
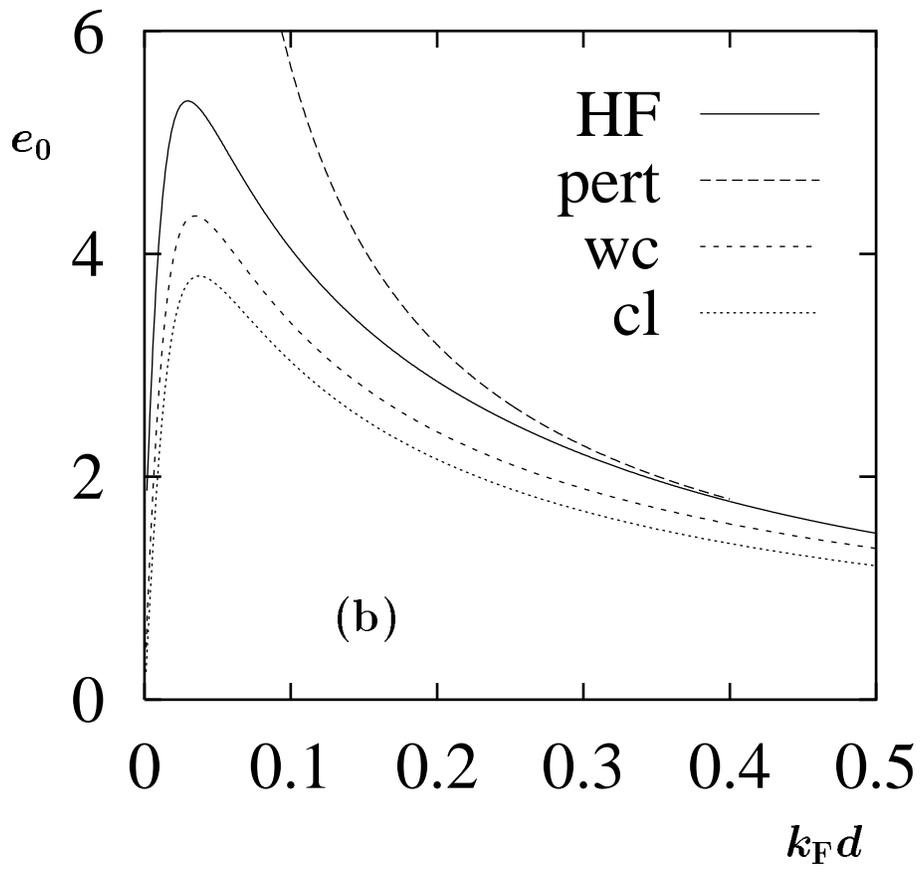
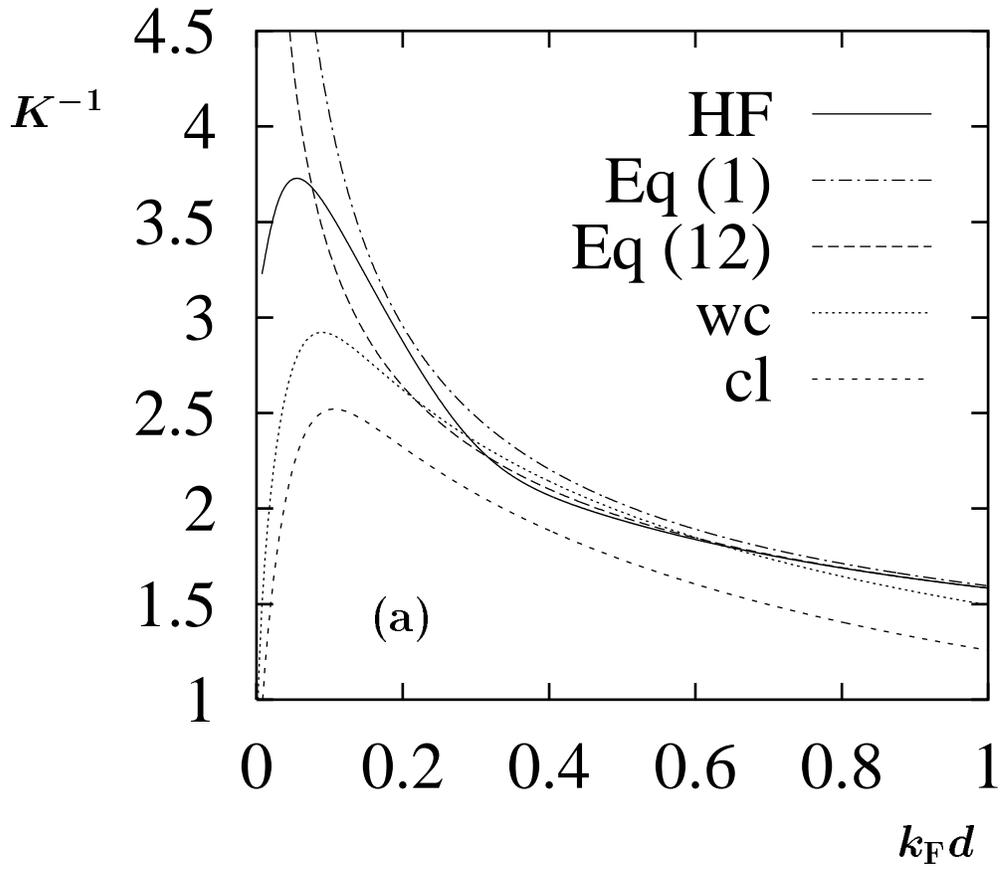
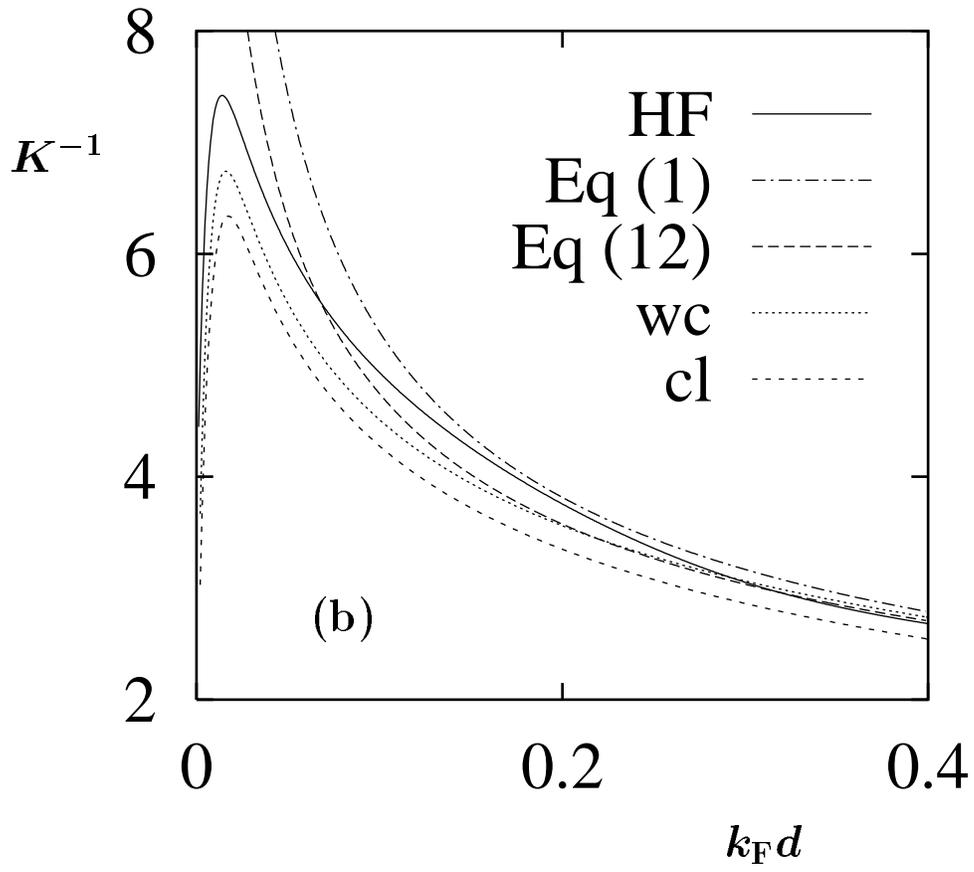
\end{document}